\documentclass[11pt,leqno]{article}
\usepackage{graphicx}
\usepackage{amssymb}
\usepackage{amsmath}
\usepackage{pifont}
\usepackage{subfig}
\usepackage{tabularx}
\usepackage{multirow}
\usepackage{multicol}
\usepackage{eepic,epic}
\usepackage{epsfig}
\usepackage{color}
\usepackage{mathptmx}
\usepackage{amsthm}
\usepackage{nicefrac}
\usepackage{booktabs}
\usepackage{todonotes}
\usepackage{tikz}
\usetikzlibrary{positioning}
\usetikzlibrary{calc}
\usepackage{enumerate}
\usepackage{rotating}
\usepackage[linesnumbered,vlined,ruled]{algorithm2e}
\input epsf %

\usepackage{url}
\definecolor{Purple}{RGB}{204, 0, 255}

\begin{document}

\title{Data Fusion Challenges Privacy: What Can Privacy Regulation Do?}

\author{ G\'{a}bor Erd\'{e}lyi \\
        University of Canterbury\\
       School of Mathematics and Statistics\\
       Christchurch, New Zealand\\
       \and
        Olivia J. Erd\'elyi  \\
        University of Canterbury\\
       School of Law\\ 
       Christchurch, New Zealand \\
       University of Bonn\\
       Center for Science and Thought\\
       Bonn, Germany\\
       \and
        Andreas W. Kempa-Liehr \\
        University of Auckland\\
        Department of Engineering Science\\
        Auckland, New Zealand}

\date{January 17, 2023}

\maketitle

\begin{abstract}
This paper focuses on some shortcomings in current privacy and data protection regulations' ability to adequately address the ramifications of some AI-driven data processing practices, in particular where data sets are combined and processed by AI systems. 
The aim of this paper is twofold: We introduce a mathematical model, which demonstrates the exploitation of complex hidden correlations in seemingly uncorrelated variables through the merging of data sets. 
Secondly, we raise attention to two regulatory anomalies related to two fundamental assumptions underlying traditional privacy and data protection approaches: \emph{Assumption 1: Only Personally Identifiable Information (PII) and Personal Data (PD) require privacy protection}. Privacy and data protection regulations are only triggered with respect to PII/PD, but not anonymous data. This is not only problematic because determining whether data falls in the former or latter category is no longer straightforward, but also because privacy risks associated with data processing may exist whether or not an individual can be identified. \emph{Assumption 2: Given sufficient information provided in a transparent and understandable manner, individuals are able to adequately assess the privacy implications of their actions and protect their privacy interests}. 
Through a review of interdisciplinary literature, we show that this assumption corresponds to the current societal consensus on privacy protection. However, relying on human privacy expectations fails to address some important privacy threats, because those expectations are increasingly at odds with the actual privacy implications of data processing practices, as most people lack the necessary technical literacy to understand the sophisticated technologies at play, not to mention correctly assess their privacy implications. To tackle these anomalies, we recommend regulatory reform in two directions: (1) Abolishing the distinction between personal and anonymized data for the purposes of triggering the application of privacy and data protection regulations and (2) developing methods to prioritize regulatory intervention based on the level of privacy risk posed by individual data processing actions. We are concerned with \emph{regulatory} anomalies and potential solutions, rather than their implementation in specific \emph{legal} regimes. This is an interdisciplinary paper that intends to build a bridge and facilitate coordination between the various communities involved in privacy research and policymaking. 
\end{abstract}

\section{Introduction}\label{Intro}

A growing body of literature as well as 
intense civil and political debates draw attention to controversial information and communication technology (ICT) practices and their far-reaching impact on human societies. Privacy and data protection and what we can do to safeguard them is at center stage in those discussions. Concerns 
reflect a general sensation that existing privacy regulations have yet again become outdated and unable to protect us from the perils of novel ICTs, see the work by Nissenbaum~\cite{HN2018} for a good overview. Yet, privacy---first mentioned in the landmark US Supreme Court case Olmstead v. United States~\cite{OvUS1928}---has always been a fluid and elusive concept, making the determination of what and how exactly we aim to protect that much harder.

In part, this is because conceptualizations of privacy differ across the various scientific disciplines concerned with it and in privacy regulations, but limited communication between these communities stands in the way of synthesizing their insights into a comprehensive privacy notion. Accordingly---as outlined in more detail in Section~\ref{RW}---protecting privacy may mean many things from respecting certain social contexts to correcting the power imbalance or appropriately defining boundaries between the individual and society, to protecting individuals' autonomy. All of these are heavily ethically loaded questions with differing interpretations across cultures, which, together, 
are important drivers of privacy laws and policies at any given time. Social science literature as well as legal and policy debates are mostly concerned with these often pronounced disciplinary, ethical, legal, and cultural differences, attempting to achieve compromises that are acceptable for a given group of stakeholders---be it a single or a group of countries, or just specific communities.     

In this contribution, we focus on a key similarity between different approaches to protect privacy: They all center around human \emph{expectations} as determinants of privacy norms. Human expectations are a natural cornerstone of privacy scholarship, laws, policies, computer science, and engineering. In light of the increasing ubiquity of artificial intelligence (AI) technologies and other modern data processing practices, these expectations are challenged.  The reason is that due to advances in computing capabilities and the widespread availability of large data sets, data processing practices have reached a level of complexity and sophistication that is no longer accessible without some technical training. Here, \emph{AI systems} are understood as defined by the Organisation for Economic Co-operation and Development (OECD)~\cite{OECD2019}
as 
\begin{quote}
a machine-based system that can, for a given set of human-defined objectives, make predictions, recommendations, or decisions influencing real or virtual environments. AI systems are designed to operate with varying levels of autonomy.
\end{quote}
Consequently, the vast majority of the population lacks the necessary knowledge to understand their workings, let alone anticipate their privacy implications. Hence, naive reliance on privacy expectations would lead to vulnerabilities in our privacy frameworks due to a discrepancy between the \emph{expected} and \emph{actual} privacy implications of AI-driven data processing practices. In our increasingly automated, digital world, privacy regulation still needs to consider individuals' privacy expectations, but needs to be supported by education and careful regulation, especially where technical means of data protection are used~\cite{KoopsLeenes2013_Privacy}.

To support our argument that human privacy expectations are, in and of themselves, unreliable privacy safeguards and to highlight that data processing does not only entail privacy risks when PII/PD is processed---we provide a simplified mathematical model for demonstrating potential consequences of combining uninformative data sets. In addition, we use Cambridge Analytica's successful interference with the 2016 US elections as an example to show that \emph{there exist} 
AI-driven data processing practices that challenge two fundamental assumptions underlying traditional privacy and data protection approaches: 
(1) Only PII/PD requires privacy protection and (2) given sufficient information provided in a transparent and accessible manner, individuals are able to  effectively protect their privacy interests. 
The mathematical model and Cambridge Analytica's modelling are, of course, by no means the only examples to illustrate the problems at hand, many other similar examples are conceivable across diverse domains. In fact, privacy economics literature asserts that individuals are virtually never completely aware of the privacy risks of their actions, and hence seldom able to give their \emph{fully informed} consent to market interactions involving their data~\cite{ATW2016}. 
We chose the Cambridge Analytica example because it is well-known and has prompted a considerable amount of jurisprudence leading to significant fines and sanctions. Yet, those \emph{legal} solutions do not address the \emph{regulatory} anomalies we discuss in this paper. 

Based on our analysis, we put forward two
recommendations for regulatory reform: 
\begin{enumerate}
    \item Due to some data processing methods---most notably the increasingly standard practice of combining data sets and processing them with AI systems---anonymity has become a fluid, dynamically evolving property of data sets. This arguably makes the traditional distinction between personal and anonymized data for the purposes of determining the scope of application of privacy and data protection regulation somewhat meaningless. In any case, it is cumbersome to implement in practice, as safeguards must work in a dynamic manner engaging and disengaging depending on whether the data is personal or anonymous at a given juncture. Also, individuals are exposed to privacy risks and may suffer harms even if they relinquish their data in anonymized form, or---as we will show---their data did not even contribute to the creation of the AI model by which it is later processed. Against this background, we propose that this distinction be abolished such that privacy protection applies to all data at all times. 
    
    \item Risk-based regulation is currently a prevailing feature of AI, privacy, and data protection regulation in both Europe and beyond~\cite{TCJ2021,RG2020,RG2021}.
Prominent examples employing a risk-based approach include the
General Data Protection Regulation~\cite{GDPR} and 
the proposed Artificial Intelligence Act ~\cite{AIA} and
ePrivacy Regulation~\cite{ePrivacyReg} in the EU.
Further examples are the ongoing work of the OECD Network of Experts on AI (ONE AI), which is implementing the OECD Principles on AI~\cite{OECD2019}, the Feasibility Study of the Council of Europe's Ad-Hoc Committee on Artificial Intelligence (CAHAI)~\cite{CAHAI2020}, and existing and ongoing standardization efforts of both European and international standardization bodies, most notably the European Committee for Standardization (CEN), the European Committee for Electrotechnical Standardization (CENELEC), the International Standardization Organization (ISO), the International Electrotechnical Commission (IEC), and the Institute of Electrical and Electronics Engineers (IEEE). In the spirit of risk-based regulation, we recommend prioritizing regulatory intervention based on the degree of privacy risk posed by individual data processing actions. 
\end{enumerate}

Throughout the paper, we use both technical and legal terms in a manner accessible for readers outside of the relevant community. We also highlight linkages between technical, legal, and other disciplinary notions relevant for addressing the privacy issues raised in the paper to facilitate interaction between various disciplines as well as the technical and policymaking communities for the purpose of solving these problems. 

The paper is structured as follows: Section~\ref{RW} reviews existing privacy concepts and provides a brief introduction to privacy regulation. Section~\ref{Model} highlights the technical side of our regulatory problems by presenting a simplified mathematical model of combining data sets. Section~\ref{Example} traces relevant aspects of the Cambridge Analytica scandal as a real-life example for how this theoretic model works in practice. Section~\ref{Recommendation} presents our recommendations for regulatory reform, and Section~\ref{Conc} concludes.

\section{Preliminaries}
\label{RW}

Privacy and data protection regulations may define which actions with privacy implications are permissible and under what circumstances, but the value judgments underlying those rules are determined by society's understanding of privacy. Rules, whose content is not in line with societal values, are unlikely to be accepted as legitimate by society. Therefore, when designing regulations, it is important to duly consider the diverging privacy conceptions that exist across society---as opposed to just relying on the law---to ensure that a widely agreed-upon privacy notion is used, regulatory objectives and methods are optimally set, in order to increase the likelihood of acceptance of the newly created norms~\cite{BCL2012,EG2022}.

This section gives an overview of existing privacy concepts across various disciplines, and familiarizes the reader with key sources of privacy and data protection regulation, relevant regulatory terminology, and the relationship between privacy and data protection. 

\subsection{Notions of Privacy}

\emph{Psychology} approaches privacy from the individual's perspective, emphasizing that it is a context-dependent, subjective feeling, which serves a self-protective purpose by controlling access to the self~\cite{AB1964,IA1975}.

\emph{Sociology} adds a macro layer to the inquiry, noting that privacy is central to building and maintaining trust both among individuals and towards social institutions. Hence, it plays a crucial role in structuring interpersonal relationships, and establishing and maintaining orderly and just societies~\cite{ACCH2017}. 

\emph{Philosophy} tackles a wide field of fundamental dilemmas related to privacy. Many of these discourses have been dominated by a long-standing deadlock between Kantian and utilitarian thinking, until Nissenbaum's doctrine of contextual integrity widely conquered both academic and policy circles~\cite{HN2009,HN2018,JR2019,PBR2012}. Conversely to classical ethical theories' endeavor to establish universally applicable behavioral guidelines, Nissenbaum's analytical framework stresses that social space is not uniform. Rather, it is composed of distinct spheres, in which nuanced, context-relative informational norms govern the flow of information. Informational norms mirror dynamically evolving privacy expectations, which differ widely across societies and define appropriate behavior in different social contexts. Contextual integrity is given as long as information flows conform with privacy expectations, but entrenched informational norms sometimes need to be reassessed and adapted in the face of changing circumstances.   

\emph{Economics}---in particular \emph{information} and \emph{privacy economics}---takes a more formal approach to privacy, investigating 
the tradeoffs associated with balancing and navigating the boundaries between public and private spheres: The decisions individuals make and the benefits and costs of those decisions for both individual agents and society. The informational dimension of privacy---the tradeoffs involved in protecting or sharing PD---is central to that inquiry~\cite{ATW2016}.

In \emph{computer science}, a recent line of work  introduced \emph{differential privacy}, a formal mathematical framework for analyzing and making data available in a privacy-preserving way~\cite{Din-Nis-Diff-Priv,Dwo-Nis-Diff-Priv,Dwork-diff-priv}. The key point of differential privacy is that a person does not necessarily have to share their PD to suffer privacy loss. Take the hypothetical example of a monthly company bulletin, which publishes the total number of employees and the number of employees who earn more than $\$500,000$ a year. Let us assume that employee $E$ leaves the company and there is no new hiring. The following month, the statistics show a decrease by one in both the number of total employees and that of high-income employees, leading to an immediate privacy loss of $E$ regarding their income category. Admittedly, this is a very constructed example but it shows that the analysis and/or release of a certain data set can have an impact on a person's privacy even if their PD is not included in that data set. In order to avoid such unintentional privacy losses, in differential privacy settings, random noise is introduced to the collected data to enable its privacy-preserving analysis and release. The so called \emph{privacy loss parameter} $\epsilon$ measures the deviation in privacy loss between a person's opt-out scenario (i.e., where the person's private information is not included in the data) and the execution of a differential privacy analysis (i.e., where the person's private information is in the data). The value of $\epsilon$  
is usually very small, meaning that including a person's private information into a data set (e.g., via a questionnaire) will not result in a significantly larger privacy loss compared to the opt-out scenario. Despite the limited setting in which \emph{differential privacy} is applied, it is an important first step in thinking about privacy in formal terms. We particularly appreciate the non-binary nature of the privacy notion: The idea of associating a privacy loss with each specific data processing step 
and assuming incremental increases in the individual's risk of exposure every time their data is processed. 

Note that the current stance in privacy and data protection regulation---which we introduce in Section~\ref{sec:privacy_basics}---seems to borrow bits and pieces from all the above mentioned disciplines, except computer science. 
We constructively challenge the weight policymakers attribute to human expectations in shaping privacy regulations.

\subsection{Privacy Regulation Basics}
\label{sec:privacy_basics}

Privacy regulators are driven by a traditional liberal view of privacy, which centers on defending individual rights against the collective and sees privacy's primary purpose in protecting individual autonomy~\cite{WB1890,AW1967,SB1984,DM2018}. Accordingly, privacy laws are characterized by a public private dichotomy and are typically only applicable to safeguard individuals' reasonable expectations of privacy over publicly not available information from unjustified intrusions. 

This understanding is reflected in leading, internationally influential hard and soft legal sources of privacy 
regulation: Recitals 47 and 50 of the European General Data Protection Regulation (GDPR)~\cite{GDPR}---the currently toughest privacy regulation in the world, which, for better or worse, has arguably become the de-facto global standard and a core benchmark for privacy regulators across the globe---both require due consideration of the reasonable expectations of identified or identifiable natural persons whose PD is processed when determining the legality of particular processing acts. Such natural persons are referred to as \emph{data subjects} in the GDPR and the Organisation for Economic Co-Operation and Development's (OECD) Privacy Framework~\cite{OECD2013}, \emph{personal information controller} in the Asia-Pacific Economic Cooperation's (APEC) Privacy Framework~\cite{APEC2015}, and \emph{PII principal} in relevant international standards, most notably the ISO/IEC 27xxx standard family on information security, see for example 
International Organization for Standardization~\cite{ISO27001,ISO27002,ISO27018,ISO27701}. The \emph{Respect for Context Principle} in the US Privacy Bill of Rights~\cite{PBR2012} expressly recognizes consumers' right to expect that their PD will only be used in ways that are consistent with the context in which it was provided. The definitions of \emph{PD}, \emph{personal information}, and \emph{PII} are also indicative of a regulatory objective to safeguard individual autonomy, by protecting information that may enable the direct or indirect identification of natural persons. Article 4 (1) of the GDPR and Article 1 (b) of the OECD Privacy Framework defines PD essentially as any information relating to a data subject, irrespective of whether it allows for direct or indirect identification of the data subject. The APEC Privacy Framework and the ISO/IEC 27xxx standards  use a different terminology---personal information (Article 9 APEC) and PII (section 3 of the standards), respectively---but the same or very similar definitions. 
Finally, the notion of consent---which forms one of the cornerstones of contemporary privacy protection---is also based on the assumption that, given sufficient information, individuals are able to understand the privacy implications of their actions---i.e., their reasonable privacy expectations are in line with reality---so that they can effectively safeguard their individual autonomy by granting or refusing consent.  

So far, we have used the terms privacy and data protection without any explanation, and noted an apparent terminological diversity regarding key privacy concepts we introduced. This diversity begs the question of whether this is mere semantics or, if not, how these concepts relate to each other. The various disciplines, whose privacy notions we reviewed, either do not differentiate clearly between privacy and data protection or distinguish an informational dimension within privacy, suggesting that privacy and data protection are either interchangeable or the latter is a subset of the former. This relationship seems to be supported by the fact that ISO/IEC 27701---the privacy extension to information security standards ISO/IEC 27001 and 27002, which was created to account for new privacy regulations such as the GDPR---mentions \emph{privacy} information management in its title, yet contains requirements that map to Articles 5 to 49 (except Article 43) of the GDPR (see Annex D of the standard). That said, there are some important distinctions to keep in mind, at least within the legal domain: (1) The definitions of PD and PII differ slightly across different jurisdictions. (2) In the EU, privacy and data protection have two distinct legal regimes linked to two different fundamental rights: Privacy is enshrined in Article 7 of the Charter of Fundamental Rights of the European Union (CFR)~\cite{CFR} and Article 8 (1) of the European Convention on Human Rights (ECHR)~\cite{ECHR} as the \emph{right to respect for private and family life}, while the \emph{right to data protection} is laid down in Article 8 CFR. Noting the different nature of both rights when it comes to protecting the private sphere, as well as their intertwined character and convoluted interaction with other rights, Gellert and Gutwirth~\cite{GG2013} essentially consider privacy and data protection as intersecting sets.

As mentioned earlier, legal rules are derived from shared societal values, and from a regulatory perspective it is paramount to ensure that the content of rules reflects those values, or else the rules cannot claim legitimacy. The regulatory issue we are dealing with here is a bit more nuanced: The problem is not that legal rules are not in line with societal values but that those values no longer correspond with technical reality. Hence, continued reliance on them without any adaptation will lead to failure to effectively protect privacy. Akin to the situation in Olmstead v United States~\cite{OvUS1928}, society needs to come to terms with yet another set  of novel technologies and ensure that the law evolves in order to adequately safeguard the core societal value of privacy.

We have also shown that society's values with regard to privacy are shaped by several disciplines. Apart from EU law, the above introduced disciplines and regulatory sources do not clearly distinguish between privacy and data protection. Also, this paper aims to tackle two \emph{regulatory} anomalies and propose \emph{regulatory} solutions to eliminate or at least alleviate them, rather than engaging with how those solutions should be implemented in specific \emph{legal} regimes. For these reasons, we use the terms privacy and data protection interchangeably throughout this paper. 

\section{Modeling Possible Consequences of Combining Data Sets}
\label{Model}

Let us give a mathematical model of combining two data sets to show two things: (1) This data processing method may enable data processors---humans and in particular AI systems---to draw inferences well beyond the scope of the individual data sets and use the inferred additional information to influence individuals' opinions and behavior. (2) These consequences are far from obvious for an average person without technical expertise. 

\begin{figure*}[t]
\begin{center}
\begin{tabular}{ll}
(a) Database $DB_A$ & (b) Database $DB_B$ \\
\includegraphics[width=0.49\textwidth]{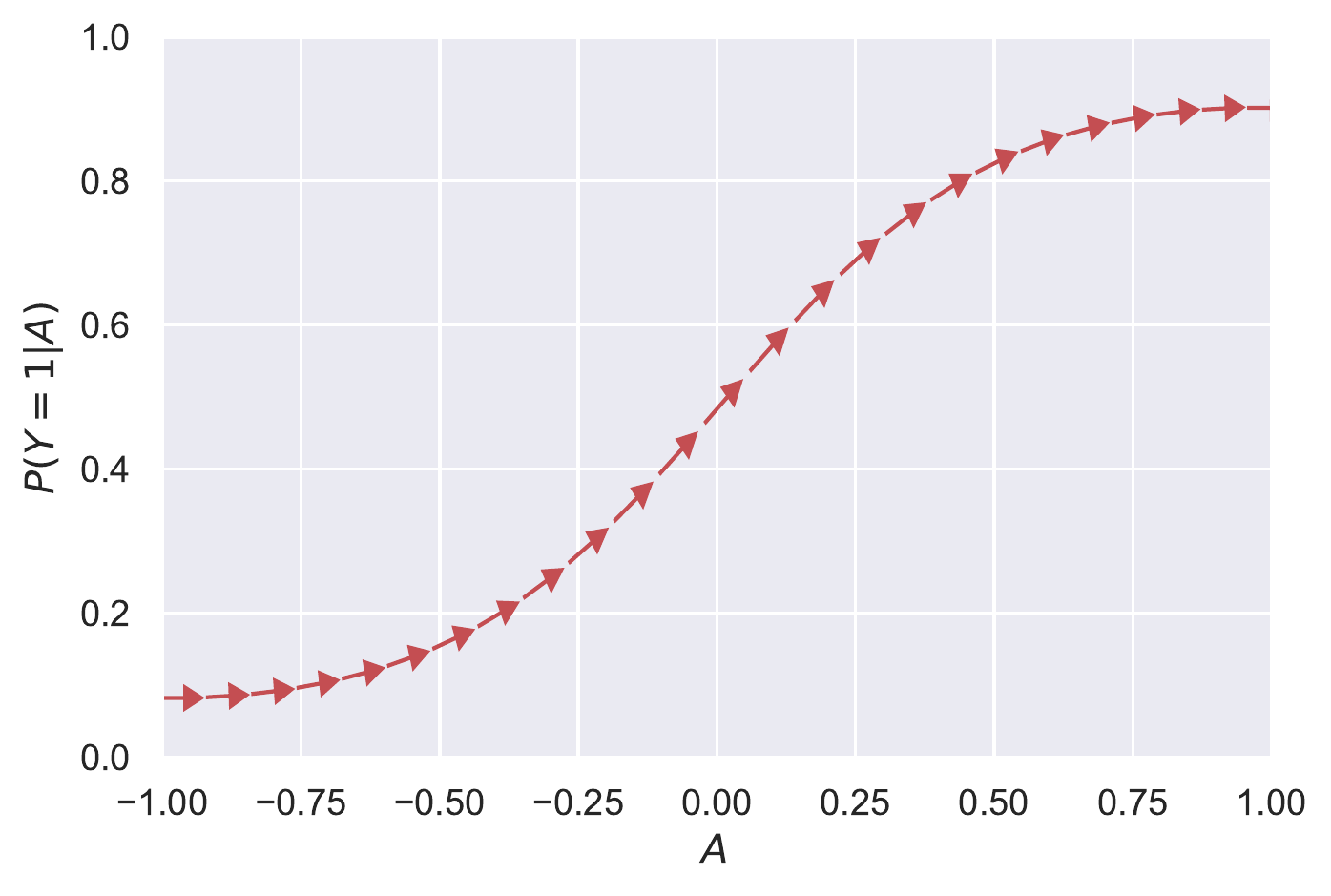} &
\includegraphics[width=0.49\textwidth]{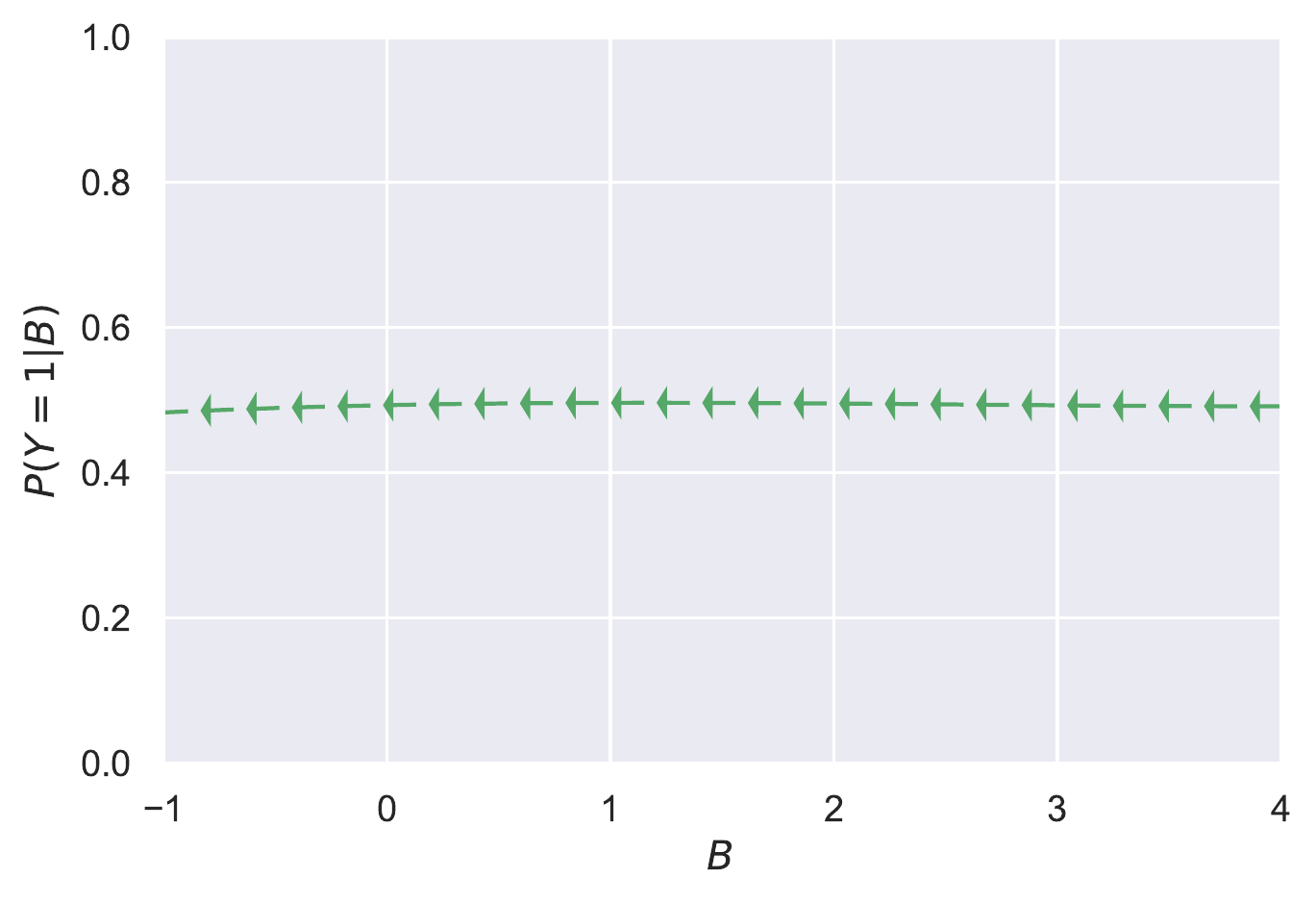} \\
(c) Databases $DB_A$ and $DB_B$ & (d) Joined databases $DB_A$ and $DB_B$\\
\includegraphics[width=0.49\textwidth]{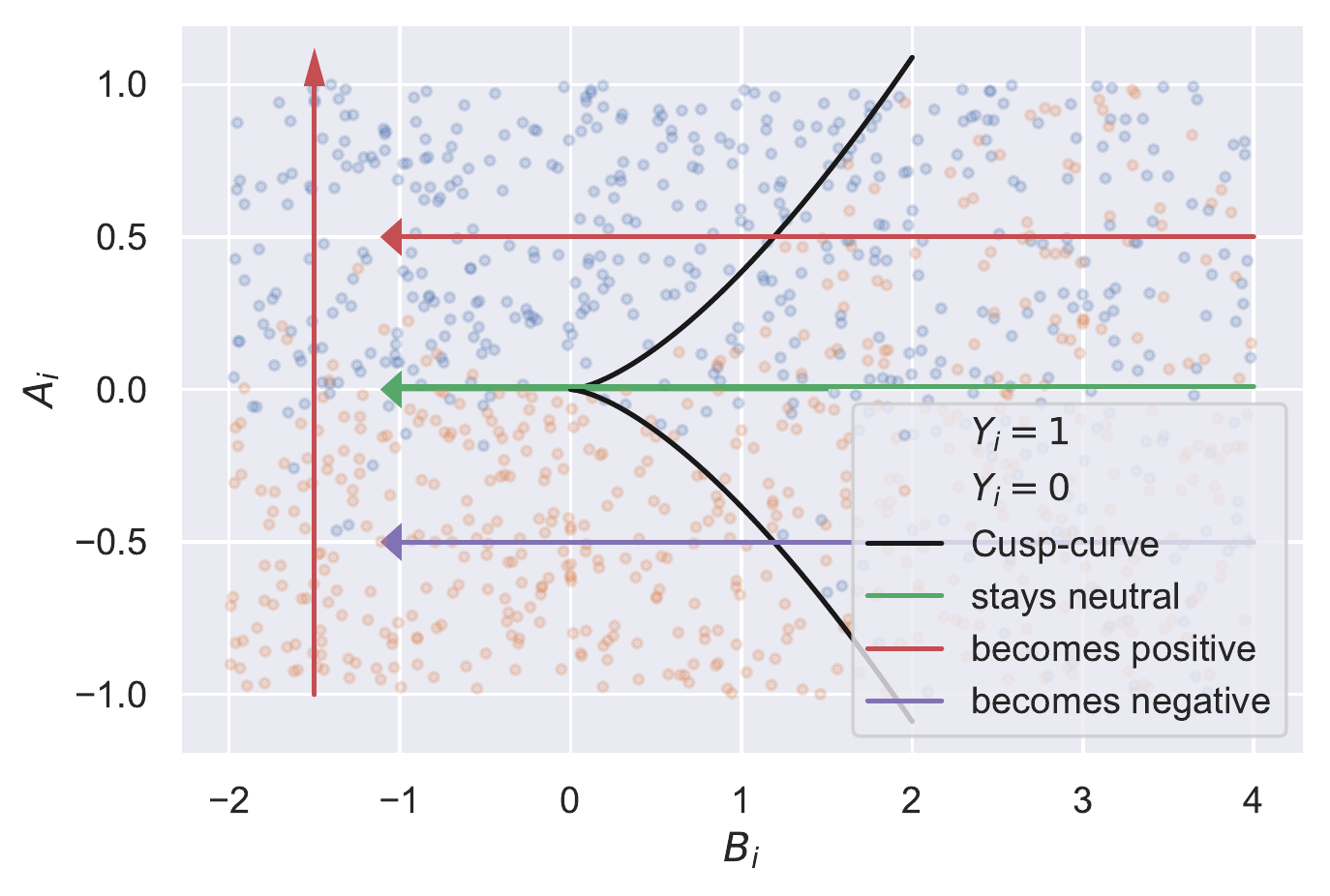} &
\includegraphics[width=0.49\textwidth]{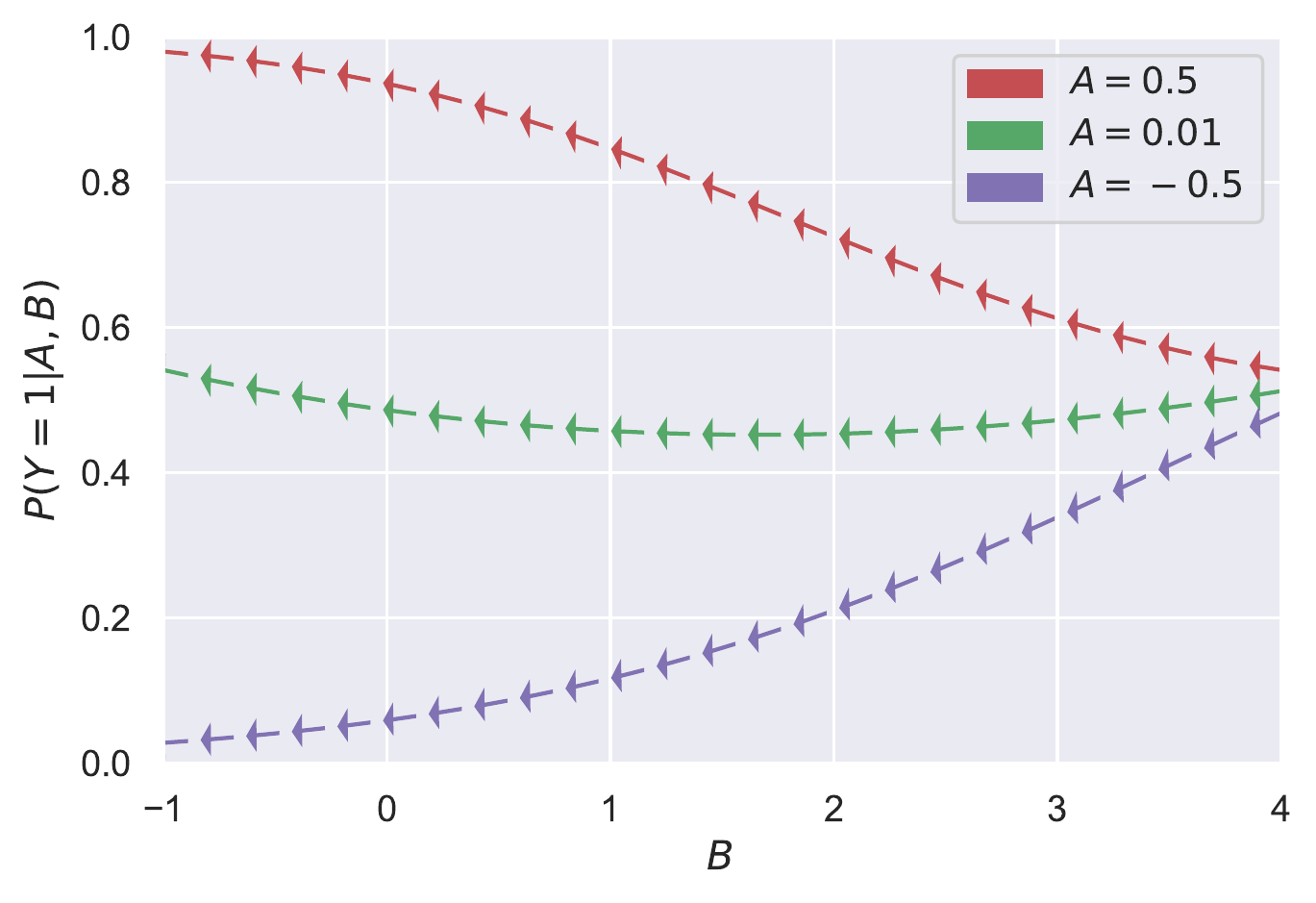} \\
\end{tabular}
\end{center}
\caption{\label{Fig:DS19ki1}
Sensitivity of conditional probabilities with respect to independent variables for system exhibiting a cusp bifurcation. (a) Probability $P(Y=1|A)$ modelled from data of $DB_A$.
(b) Probability $P(Y=1|B)$ modelled from data of $DB_B$. 
(c) Sampled data from nonlinear system exhibiting cusp bifurcation based on variables $A$ and $B$ (cf. Appendix \ref{Sec:Sampling}).
Decisions $Y_i$ are fairly unambiguous left of the cusp curve and random right of the cusp curve. 
(d) Probability $P(Y=1|A,B)$ modelled from joined databases $DB_A$ and $DB_B$. 
 Probabilities are estimated from logistic regression with L2-norm on 3rd-order polynomial features considering only one independent variable (panels (a) and (b)), or both variables (panel (d)). Arrows indicate direction of change for independent variables $A$, $B$.
}
\end{figure*}

Assume that we are dealing with two databases $DB_A$ and $DB_B$, which are located at two different internet service providers.
Database $DB_A$ includes a demographic index $A$, which is an indicator of a person's probability to vote on a specific topic $Y$ (Figure~\ref{Fig:DS19ki1}a).
For reasons of simplicity, it is assumed that the vote can only have two values: A positive ($Y=1$) and a negative ($Y=0$) choice.

Database $DB_B$ includes a social media behavior index $B$, which is designed to quantify the interaction of a user with a social media platform.
From an internal survey of the social media platform, which asks for the user's preference in the upcoming vote, someone tries to model user $i$'s vote $Y_i$ based on the corresponding index $B_i$. As it turns out, $Y_i$ and $B_i$ are statistically independent $P(Y,B)=P(Y)P(B)$ and $P(Y=1|B)\approx 50\%$ (Figure \ref{Fig:DS19ki1}b).

\begin{figure*}[t]
\begin{tabular}{ll}
(a) Bistability of latent variable $X$ & (b) Sampled votes $Y_i$\\
\includegraphics[width=0.47\textwidth]{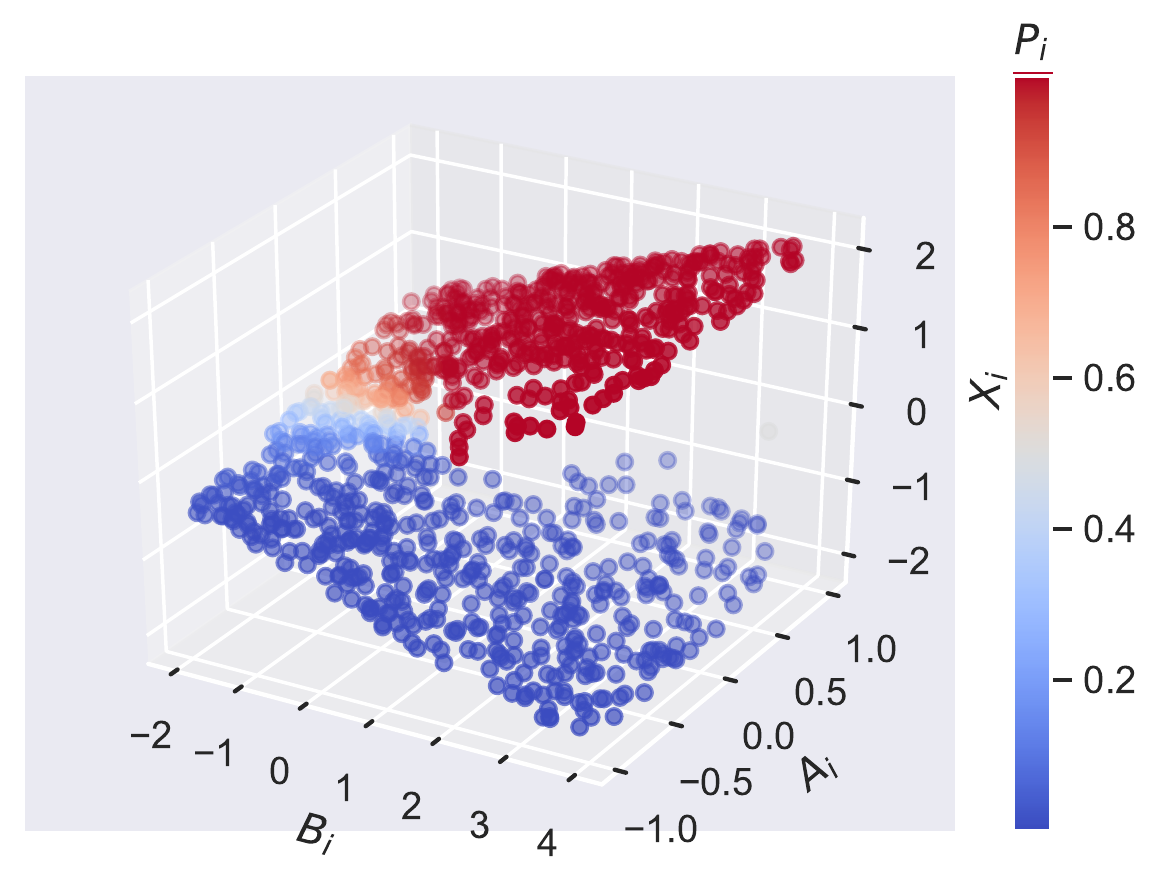} &
\includegraphics[width=0.47\textwidth]{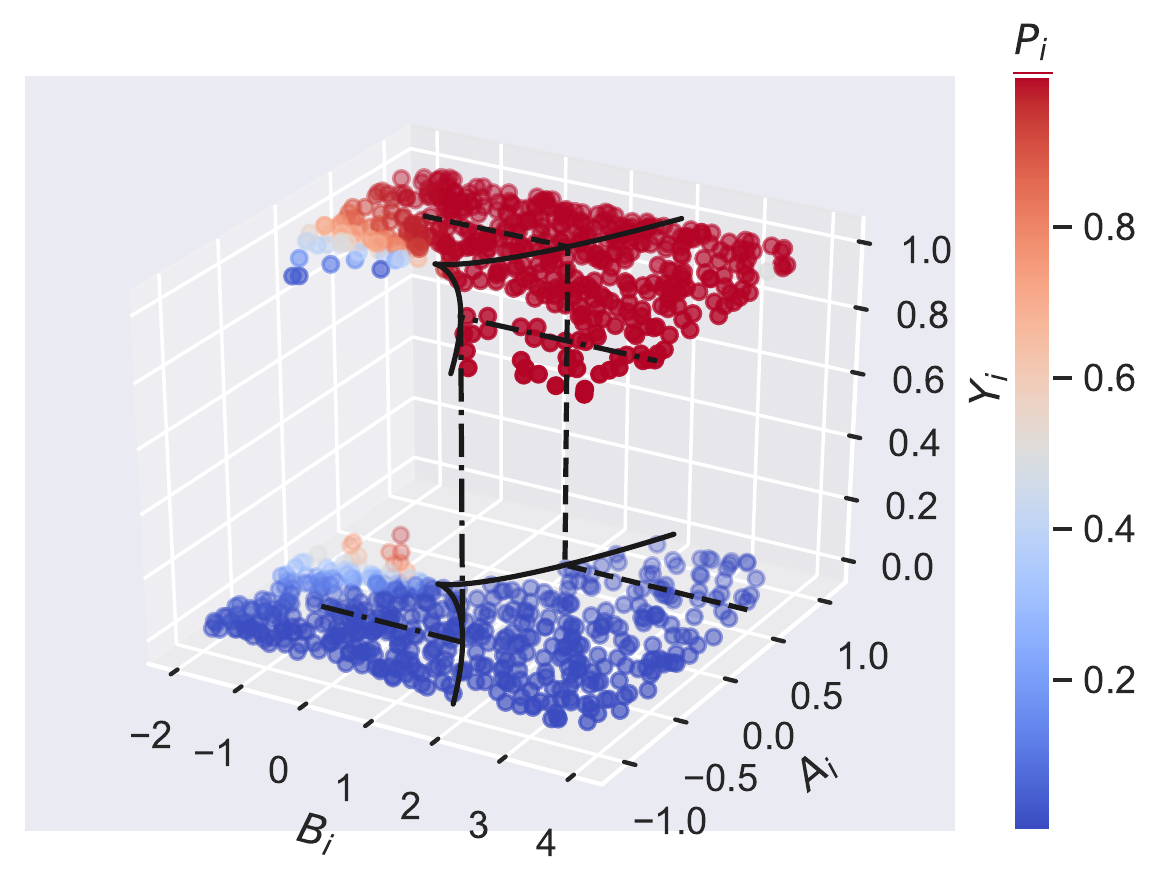} \\
\end{tabular}
\caption{\label{Fig:DS19ki0}Bistability of latent variable $X$ (a) and sampled votes (b). The sampling is described in Appendix A. 
Subfigure (b) also shows the cusp curve Eq.\eqref{Eq:cusp_curve} (solid line) and the sudden changes of voting decisions happening at $B=\frac{3}{2\sqrt[3]{2}}$ for $|A|=\frac{1}{2}$ (dashed lines). For details of the probabilistic model, refer to Appendix \ref{Sec:Sampling}.
}
\end{figure*}

However, due to the nature of this constructed problem, it is assumed that variables $A$ and $B$ are competing indicators with respect to the person's decision on the upcoming vote, such that the overall dynamics can be modeled as a nonlinear system exhibiting a cusp bifurcation \cite{Zeeman1976_catastrophe} using the Cusp function
\begin{equation}\label{Eq:CuspFunction}
	V(x;A,B) = \frac{1}{4}x^4-Ax-\frac{1}{2}Bx^2,
\end{equation}
which has two local minima for 
\begin{equation}\label{Eq:cusp_curve}
|A|\le 2\left(\frac{B}{3}\right)^{\frac{3}{2}}
\end{equation}
and one local minimum otherwise (Figure~\ref{Fig:DS19ki0}a).
The local minimum $X_i$ for a specific person $i$ with $A_i$ and $B_i$ fulfills
\begin{equation}\label{Eq:local_minimum}
\frac{\partial}{\partial x}\left.V(x;A_i,B_i)\right|_{x=X_i} = 0
\end{equation}
so that it constitutes a latent variable, from which the person's probability $P_i$ to vote $Y_i=1$ can be inferred
from a simple logistic regression model \begin{equation}\label{Eq:P_i}
P_i = P(Y_i=1) = (1+\text{e}^{-\sigma X_i})^{-1},
\end{equation}
which can be considered as a blueprint for more complex machine learning algorithms.
A sample of 1000 simulated tuples $(A_i, B_i, Y_i)$ is shown in Figure~\ref{Fig:DS19ki0}b.

Consequently, a small change of $B$ could result into a spontaneous switching of the person's vote, if this change crosses the cusp-curve 
\begin{equation}
|A|\le 2\left(\frac{B}{3}\right)^{\frac{3}{2}}=\frac{2B^\frac{3}{2}}{3\sqrt{3}}
\end{equation}
and the person's local minimum vanishes (Figure~\ref{Fig:DS19ki1}c).
From the perspective of statistical modeling, this means that joining the information $A_i$ and $B_i$ from both databases $DB_A$ and $DB_B$ allows the formulation of a model, which can predict the user specific change of opinion $P_i$ by changing their social media behavior $B_i$ (Figure~\ref{Fig:DS19ki1}d). For the details of our sampling, we refer the reader to technical Appendix~A. 

This simplified example shows that both identifying the individual(s) whose opinion(s) is/are closest to the cusp curve (and hence most easily changed) and calculating how much influence to exert upon them require complex mathematical modeling. Of course, its efficiency depends on the characteristics of the available data sets. Let us also assume that the relevant terms and conditions, to which individuals consent when providing their PD, are fully transparent on how information is processed, likely involving reference to the possibility of \emph{combining} data sets and \emph{processing} them \emph{by automated means}. Still, in the absence of at least some technical literacy, people will not realistically anticipate such far-reaching implications. We examine this claim in more detail in the next section.

\section{How It Works In Practice: Cambridge Analytica and the 2016 US Elections}
\label{Example}

Based on a study of the European Parliamentary Research Service~\cite{EPGDPR2020} and Isaak and  Han\-na~\cite{Isa_han:j:Privacy}, we now outline relevant aspects of the infamous Cambridge Analytica scandal as a real-life example, where data merging 
coupled with extensive use of machine learning (ML) 
algorithms 
has been employed to 
influence voters' behavior in the 2016 US Elections. Information on terms and conditions and other pieces of communication between the parties involved in the Cambridge Analytica exercise was not always available. 
In such instances, we assume GDPR-compliant behavior in our example to show that the problems discussed below occur even if these currently highest (or most restrictive) standards of data protection are met.    
For the same reason, we also use the GDPR as a benchmark to evaluate other aspects of the event, even though it occurred in the US and was hence beyond the GDPR's scope. 
Note, that people affected by this incident did not anticipate the privacy implications of their actions. This supports our point that the average individual---despite safeguards in privacy regulation---is unable to adequately assess the far-reaching consequences of AI-driven data processing practices.

Because the example is in a supervised learning setting and statistics, statistical learning, machine learning, and data mining  literature use multiple terms to refer to some core concepts, 
let us briefly explain the notion of supervised learning and give an overview of the choices of terminology in use. We will follow the notation of James et al.~\cite{JWHT-stat-learning}. In \emph{supervised learning} we are given a vector $X$ of $p$ \emph{predictor variables} 
(also known as input, feature, or independent variables) and a \emph{response variable} $Y$ (also known as outcome, target, or dependent variable). We are assuming that there is some kind of a relationship between $X$ and $Y$. Furthermore, we are given a \emph{training data set} of $n$ observations of the form $\{ (x_i,y_i):i=1,2,\ldots ,n\}$, where $x_i$ is a vector of $p$ predictor variables and $y_i$ is the response value for the $i^\text{th}$ observation. If $Y$ is quantitative, we have a \emph{regression problem}. If $Y$ takes values of a finite set, we have a \emph{classification problem}. In supervised learning, using the training data set, we try to understand the relationship of $X$ and $Y$ and 
how each predictor variable affects the response. With  this knowledge, we use observed predictor vectors $x$ in order to predict an unobserved $y$.  

Turning to the example, Cambridge Analytica's objective was to (1) build an ML model that can infer individuals' personality traits (response variable) 
based on their Facebook data (predictors) 
and (2) use this information to sway swing voters' preferences in the desired direction. We focus on the first part of this goal here because it is the one involving the combination of multiple data sets and their processing by ML algorithms---the data processing practices whose privacy implications we aim to analyze---and enabling the execution of the second part. What follows is a description of four privacy-relevant actions that Cambridge Analytica has taken to achieve the first part of the above outlined objective. To help highlight the connection between relevant technical and legal concepts, as well as their likely interpretation by lay people, we do not only describe the respective actions but also analyze them from a technical (\emph{technical side}) and legal (\emph{legal side})  perspective, and discuss the validity of probable \emph{privacy expectations} related to them. Note that we give a simplified account of events, only going into details relevant for our purposes.

{\bf Action 1---Creation of training set:} First, Cambridge Analytica had to create a training data set. As a first step to achieve that, they have paid roughly $300,000$ people to fill out the short version of a psychological survey\footnote{The International Personality Item Pool Representation of the NEO PI-$\mbox{R}^{TM}$ survey~\cite{NEO-PRI}.} via a survey app. \emph{The technical side}: At this point, anonymous data is collected and stored, creating the preconditions of further data processing. \emph{The legal side}: The stated purpose of data collection is research. 
Because of the anonymous nature of the survey data, its processing---\emph{collection} and \emph{storage} are forms of processing according to Article 4 (2) GDPR---falls squarely outside the  purview of the GDPR as defined in Article 2 GDPR, so no specific information or consent requirements apply. This is a point of paramount importance to which we will return later, as such requirements would signal individuals the existence of privacy risks, potentially leading them to refrain from filling out the survey. 
\emph{Privacy expectations}: From a lay person's perspective, giving up the information required by the survey has no privacy implications, as it is used for a valid purpose and---as far as participants are concerned---cannot be traced back to them. Participants' privacy \emph{expectations} (no privacy risks) are thus very different from the \emph{actual} privacy implications of filling out the survey (high privacy risks due to enabling further data processing). 

{\bf Action 2---De-anonymization of training data:} To receive payment for filling out the survey, users had to login to their Facebook account and approve access for the survey app. Using this access, the app has collected all data associated with them (including PD, links, posts, likes, and all the Facebook data of their friends). 
\emph{The technical side}: 
Several key steps with substantial privacy implications happen here. 
For one, the hitherto anonymous survey data is linked to participants' Facebook accounts and is thereby de-anonymized. This enables Cambridge Analytica to merge these two data sets---survey data revealing participants' personality traits and data contained in their Facebook accounts---providing them with their initial training data set.  
\emph{The legal side}: De-anonymization turns the thus far anonymized survey data into \emph{PD} within the meaning of Article 4 (1) GDPR, triggering the GDPR's application. We could not access the app's terms and conditions, so we assume that, complying with the GDPR, participants are informed that granting the app access to their Facebook account allows Cambridge Analytica to collect, store, and otherwise process information from those accounts. Let us also assume that participants are explicitly made aware of the possibility to \emph{combine} the two data sets and \emph{process} them \emph{by automated means}---two other forms of processing according to Article 4 (2) GDPR---and that, based on this information, they consent to the further processing of their now PD.
\emph{Privacy expectations}: Assuming participants actually read the app's terms and conditions, they presumably understand that they are providing PD stored on their Facebook account, even though it is unlikely that they realize the quantity and value of data they are giving up. However, based on the information provided in plain language in the terms and conditions, a person without technical training can arguably not anticipate the above outlined consequences of merging several data sets, nor the amount and value of information ML algorithms can extract from such combined data sets. So, it is fair to assume that the \emph{perceived} privacy risks of action 2 remain well below its \emph{actual} privacy risks. 

{\bf Action 3---Extension of training data set:} Furthermore, consumer data was acquired from other parties, whereby the identity of those parties, the exact nature of the collected data, and the terms and conditions under which the data was collected is unknown. \emph{The technical side}: We assume those additional data sets---which are combined with Cambridge Analytica's existing data sets---contain PD, seeing as this step still aims to contribute to building the training data set by further enriching it. 
\emph{The legal side}: Again, based on GDPR requirements, we assume the following: (1) Data subjects providing their PD are informed of and consent to the purpose(s) and potential form(s) of data processing. (2) This information includes reference to the possibility of processing the provided data by all the aforementioned means and by sharing it with third parties---a processing action covered by the \emph{disclosure by transmission, dissemination or otherwise making available} options of Article 4 (2) GDPR. (3) All this information is conveyed to data subjects in \emph{data controllers'/processors'}---the terms used by Article 4 (7) and (8) GDPR for referring to parties on behalf of whom PD is processed/who perform the processing---respective terms and conditions. 
\emph{Privacy expectations}: As noted in respect of privacy expectations related to action 2, we are of the opinion that even if the affected individuals are clearly and unambiguously informed about all data processing purposes and methods,
those without technical literacy are not in the position to adequately assess the aggregate privacy implications of the data processing actions involved.

{\bf Action 4---Model fitting:} Based on the collected data sets and using machine learning algorithms, Cambridge Analytica built their model with the objective to profile \emph{any} person based on their Facebook data. \emph{The technical side}: There are two elements to distinguish within this action, both posing very high privacy risks. The first one is the process of \emph{model building}, that is, the use of statistical learning methods to find the correlation between the predictors (e.g., Facebook data, consumer behavior) contained in the various data sets and the response (i.e., the personality traits associated with the predictors). Even though built based on the data of a relatively small number of individuals, this model---as all well-built models---is a very powerful tool that can subsequently be used to draw inferences from similar data on \emph{any} individual. This ability of the model enables the second element of Action 4, namely for Cambridge Analytica to \emph{engage the model} to profile any US voter based on their Facebook (and potentially other) data. 
\emph{The legal side}: Building a model and using it on data qualifies as \emph{processing data by automated means} according to Article 4 (2) GDPR. From this follows that---in respect of the survey participants and assuming that all other requirements of Article 5 GDPR are also fulfilled---all the data processing steps undertaken under Action 4 are compliant with the GDPR due to being covered by their consent to the terms and conditions of the app mentioned under Action 2, as well as to those of other data controllers/processors mentioned under Action 3. Perhaps less obviously, however, the same holds for any member of the US voter population, whose 
data is later inputted in the model, as long as Facebook's (and potentially other data controllers'/processors') terms and conditions require those individuals' consent to collecting, storing, and processing their data by sharing it with third parties, combining data sets, and processing them by automated means---which we assume here. Alternatively, insofar as data is \emph{manifestly made public by the data subject} (e.g. data in a public Facebook account), consent is not even needed according to Article 9 (2) (e) GDPR.
\emph{Privacy expectations}: The above discussion shows that---provided all other conditions of Article 5 GDPR are also fulfilled---a valid consent given to a set of sufficiently comprehensive terms and conditions at the time of first providing PD to a single data controller/processor may cover a series of privacy-relevant data processing actions, 
despite the fact that the degree to which privacy risks may actually materialize at a later time is entirely unforeseeable for the data subject. Similar privacy risks arise in an even more covert fashion if data is collected in anonymized form and hence without requiring consent. We believe it is impossible for anyone---including technical experts---to anticipate what data sets will be combined, what models will be built and used on their anonymized and/or PD, and to what purposes (which are easily formulated so as to be compliant with the purpose limitation principle laid down in Article 5 (1) (b) GDPR) 
in the future. \emph{Expected} privacy risks are, therefore, once again well below \emph{actual} privacy risks.  

Finally---turning to the second part of Cambridge Analytica's objective---after discovering voters' personality traits by running the model on their Facebook (and potentially other) data, it was a straightforward exercise 
to show them personalized advertisements with a view to actively manipulate their voting behavior in the desired direction---an undertaking that ultimately brought about the unexpected outcome of the 2016 US elections.
 
\section{Possible Regulatory Changes and the AI-Privacy Tradeoff}
\label{Recommendation}

After providing a simplified mathematical model and a real-life example of how the potentially far-reaching consequences of merging data sets and processing them by ML algorithms challenges current data protection approaches, we now put forward two recommendations for regulatory reform 
to address the above identified privacy risks. 

{\bf Recommendation 1:} \emph{Abolishing the distinction between personal and anonymized data for the purposes of determining the scope of application of data protection laws and protecting all data}: As mentioned earlier, most data protection laws and global data protection standards only come into play if PD/PII is processed. Due to the dynamic workings of these rules---they apply only as long as 
the data processed qualifies as PD/PII
---they provide relatively effective protection \emph{if and to the extent to which} data controllers/processors choose to comply with them. This \emph{will to compliance} makes a huge practical difference from a regulatory perspective, because breaches must be \emph{detected} in order to be prosecuted. Detection is, however, generally a central regulatory problem in practice due to limited regulatory resources~\cite{BCL2012}. This holds all the more true in data protection regulation in light of the enormous frequency of data processing actions. Existing solutions to enhance privacy protection, like differential privacy, do not address this problem, as their application heavily depends on data controllers'/processors' ethical judgment. The fact that existing data protection rules do not provide any privacy protection when individuals give up data in anonymized form is equally troubling. As shown by the Cambridge Analytica example, this is an important shortcoming, because this initial act of data collection enables further data processing (which may or may not be compliant with data protection laws) and hence has potentially severe privacy implications. Also, the lack of application of data protection laws gives people a false signal that giving up data in anonymized form does not entail privacy risks. We think that educating individuals about the possible privacy implications of providing anonymized data and requiring their consent to the processing of such data in a similar manner as is required for PD would induce more caution---possibly keeping people from providing information in the first place---and significantly bolster privacy protection. The recently proposed EU ePrivacy Regulation~\cite{ePrivacyReg} seems to go in the same direction, in that it provides protection for \emph{electronic communications data} even if it does not qualify as PD, recognizing that the processing of such data may also give rise to privacy risks.

{\bf Recommendation 2:} \emph{Applying risk-based regulatory principles to prioritize regulatory intervention based on the level of privacy risk posed by individual data processing actions}: As noted in the introduction, existing and forthcoming European and key international regulatory instruments on data protection and AI adopt a risk-based approach to prioritize regulatory intervention. There is also a clear trend to align international, regional, and domestic policies and standards as far as possible to create an internationally consistent, interoperable regulatory framework in these domains. Some solutions laid down in these instruments could be useful starting points for developing 
approaches to target individual data processing actions based on the level of privacy risk they pose. We note that implementing such an approach is not without challenges: For one, setting up a risk assessment framework for data processing actions is far from easy, especially given the continuous evolution of data processing methods. The idea of attributing a given privacy loss to individual data processing actions mentioned in the context of differential privacy may be helpful here. 
Second, detecting relevant processing actions and requiring data subjects' consent for each and every one of them is a serious challenge that is unlikely to be feasible without at least some level of automation. That said, the practical implementation of any automation---which in itself would involve a loss of privacy---is a formidable task that should not be underestimated. Ideas like engaging an algorithmic agent to help preserve individuals' human agency~\cite{IEEE2019} may, at first sight, seem promising to crack this problem. Depending on how they are implemented, however, they may amount to a surveillance nightmare rather than privacy protection: 
Any centralized database, through which algorithmic data processing systems can interact and access individuals' preferences, is vulnerable to abuse. Decentralized solutions may alleviate some of the concerns raised by centralized settings, but they bring a different set of problems. This very simple example is just to illustrate the daunting challenges of implementing any form of automation. Exploring various alternatives is beyond the scope of this paper.    

As a final remark, we believe society needs to come to terms with the fact that there is a tradeoff between the use of AI technologies and privacy. There is a clear rise in demand for the comfort and benefits AI may provide in many domains, such as healthcare, finance, and transportation, to name a few, as well as for ever more efficient personalized services like movie recommender systems and customer services more generally. While we should strive to provide state-of-the-art privacy protection at all times, we have to accept that the use of AI technologies inevitably involves privacy losses compared to the pre-AI age. So the real question is whether and to what extent we are willing to give up our privacy in exchange for reaping the benefits of AI. This is a decision for society to make, whereby the aim should be to find a socially optimal balance, mindful of the negative economic consequences of overly restrictive privacy policies~\cite{ITIF2021}. Given the transnational nature of privacy protection, international and multi-stakeholder coordination will be indispensable. Arguably, there must be checks and balances in place to keep powerful actors in check---the blame is most commonly assigned to Big Tech, although governments and other actors also tend to wield significant power. That said, it is important to remember this tradeoff when determining whether and to what extent these parties overstep their boundaries while serving a very real demand. 

\section{Conclusion}\label{Conc}

This paper proposed regulatory reform to address some problems stemming from AI-driven data processing practices, in particular from combining data sets---a well-known but to date insufficiently addressed issue, which has recently seen a sharp increase in significance due to the abundance of available data and advances in computing and AI technologies. Solving these problems will require policy efforts involving transnational, interdisciplinary and multi-stakeholder collaboration, as well as educating people about AI technologies and the privacy challenges they present, all of which, although much-discussed, remain tricky in practice. To highlight and facilitate the navigation of technical issues for policymakers, we not only presented an abstract mathematical model of combining data sets but also analyzed Cambridge Analytica's interference with the 2016 US elections---a real-life example illustrating how powerful this data processing method coupled with the use of AI systems can be---from a technical, legal, and lay perspective. Our recommendations are in line with ongoing and/or impending policy and standard setting work in the EU, OECD ONE AI, European and international standard setting organizations, and other bodies. We referred to publicly available components of these work streams throughout the paper.

\begin{appendix}
\section{Sampling binary choice from bistable system}
\label{Sec:Sampling}
In order to simulate a system exhibiting the sudden change of voting behavior, we are sampling $i=1,\ldots,1000$ data points from
\begin{eqnarray}
A_i   & \sim & \text{Uniform}(-1,1), \\
B_i   & \sim & \text{Uniform}(-2,4),\text{ and}\\
X_{0,i} & \sim & \text{Uniform}(-1,1).
\end{eqnarray}
In order to find combinations of $A_i$, $B_i$, and $X_i$, which fulfill Equation~\eqref{Eq:CuspFunction}, samples $X_{0,i}$ are used as initial guess for finding local minima of
\begin{equation}
X_i = \arg\min_{x} V(x;A_i,B_i)
\end{equation}
using the Nelder-Mead simplex algorithm \cite{NelderMead1965_minimization} as implemented by the Python module SciPy Ver. 1.2.1 \cite{scipy}.
The resulting samples $(A_i, B_i, X_i)$ fulfill Equation~(\ref{Eq:local_minimum})
and are depicted in Figure~\ref{Fig:DS19ki0}a.

In order to sample a classification response 
$Y_i$ from these data, the stationary points $X_i$ are mapped to probability $P_i$ by the logistic function
\begin{equation}
P_i = (1+\text{e}^{-\sigma X_i})^{-1}
\end{equation}
with $\sigma=10$. The classification response 
$Y_i$ is sampled from
\begin{equation}
Y_i \sim \text{Bernoulli}(P_i).
\end{equation}
The resulting samples $(A_i, B_i, Y_i)$ are shown in Figure~\ref{Fig:DS19ki0}b.
\end{appendix}

{\small 
\bibliography{main}

\newcommand{\etalchar}[1]{$^{#1}$}
\begin{thebibliography}{MKGCM14}

\bibitem[ACCH17]{ACCH2017}
Denise Anthony, Celeste Campos-Castillo, and Christine Horne.
\newblock Toward a sociology of privacy.
\newblock {\em Annual Review of Sociology}, 43:249--269, July 2017.

\bibitem[{Ad }20]{CAHAI2020}
{Ad Hoc Committee on Artificial Intelligence}.
\newblock {Feasibility Study}, 2020.
\newblock CAHAI(2020)23.

\bibitem[Alt75]{IA1975}
Irwin Altman.
\newblock {\em The Environment and Social Behavior: Privacy, Personal Space,
  Territory, Crowding}.
\newblock Brooks/Cole Pub. Co., first edition, 1975.

\bibitem[{Asi}15]{APEC2015}
{Asia-Pacific Economic Cooperation}.
\newblock {APEC Privacy Framework}, 2015.
\newblock
  https://www.apec.org/docs/default-source/publications/2017/8/apec-privacy-framework-(2015)/217\_ecsg\_2015-apec-privacy-framework.pdf?sfvrsn=1fe93b6b\_1.

\bibitem[ATW16]{ATW2016}
Alessandro Acquisti, Curtis Taylor, and Liad Wagman.
\newblock The economics of privacy.
\newblock {\em Journal of economic Literature}, 54(2):442--492, 2016.

\bibitem[Bat64]{AB1964}
Alan~P. Bates.
\newblock Privacy---a useful concept?
\newblock {\em Social Forces}, 42(4):429--434, May 1964.

\bibitem[BCL12]{BCL2012}
Robert Baldwin, Martin Cave, and Martin Lodge.
\newblock {\em {Understanding Regulation Understanding Regulation: Theory,
  Strategy, and Practice}}.
\newblock Oxford University Press, 2012.

\bibitem[Ben84]{SB1984}
Stanley~I. Benn.
\newblock Privacy, freedom, and respect for persons.
\newblock In Ferdinand~D. Schoeman, editor, {\em Philosophical Dimensions of
  Privacy: An Anthology}, pages 223--244. Cambridge University Press, 1984.

\bibitem[CD21]{ITIF2021}
Nigel Cory and Luke Dascoli.
\newblock {How Barriers to Cross-Border Data Flows Are Spreading Globally, What
  They Cost, and How to Address Them}, 2021.
\newblock Information Technology \& Innovation Foundation.

\bibitem[CJ21]{TCJ2021}
Tim Clement-Jones.
\newblock {How the OECD’s AI System Classification Work Added to a Year of
  Progress in AI Governance}, 2021.

\bibitem[{Cou}50]{ECHR}
{Council of Europe}.
\newblock {European Convention on Human Rights}, 1950.

\bibitem[DMNS06]{Dwork-diff-priv}
Cynthia Dwork, Frank McSherry, Kobbi Nissim, and Adam Smith.
\newblock Calibrating noise to sensitivity in private data analysis.
\newblock In Shai Halevi and Tal Rabin, editors, {\em Theory of Cryptography},
  pages 265--284. Springer, 2006.

\bibitem[DN03]{Din-Nis-Diff-Priv}
Irit Dinur and Kobbi Nissim.
\newblock Revealing information while preserving privacy.
\newblock In {\em Proceedings of the Twenty-Second ACM SIGMOD-SIGACT-SIGART
  Symposium on Principles of Database Systems}, page 202–210. Association for
  Computing Machinery, 2003.

\bibitem[DN04]{Dwo-Nis-Diff-Priv}
Cynthia Dwork and Kobbi Nissim.
\newblock Privacy-preserving datamining on vertically partitioned databases.
\newblock In {\em Advances in Cryptology -- CRYPTO 2004}, pages 528--544.
  Springer, 2004.

\bibitem[EG22]{EG2022}
Olivia~J Erd{\'e}lyi and Judy Goldsmith.
\newblock Regulating artificial intelligence: Proposal for a global solution.
\newblock {\em Government Information Quarterly}, 2022.
\newblock Availabla online 22 July 2022, 101748.

\bibitem[{Eur}12]{CFR}
{European Parliament, the Council and the Commission}.
\newblock {Charter of Fundamental Rights of the European Union}, 2012.
\newblock 2012/C 326/02.

\bibitem[{Eur}16]{GDPR}
{European Parliament and European Council}.
\newblock {Regulation (EU) 2016/679 of the European Parliament and of the
  Council of 27 April 2016 on the protection of natural persons with regard to
  the processing of personal data and on the free movement of such data, and
  repealing Directive 95/46/EC (General Data Protection Regulation)}, 2016.
\newblock https://eur-lex.europa.eu/eli/reg/2016/679/oj.

\bibitem[{Eur}17]{ePrivacyReg}
{European Parliament and European Council}.
\newblock {Proposal for a Regulation of the European Parliament and of the
  Council concerning the respect for private life and the protection of
  personal data in electronic communications and repealing Directive 2002/58/EC
  (Regulation on Privacy and Electronic Communications) and Amending Certain
  Union Legislative Acts}, 2017.
\newblock COM/2017/010 final.

\bibitem[{Eur}20]{EPGDPR2020}
{European Parliament and European Council}.
\newblock {The impact of the General Data Protection Regulation (GDPR) on
  Artificial Intelligence}, 2020.
\newblock
  https://www.europarl.europa.eu/RegData/etudes/STUD/2020/641530/EPRS\_STU(2020)641530\_EN.pdf.

\bibitem[{Eur}21]{AIA}
{European Parliament and European Council}.
\newblock {Proposal for a Regulation of the European Parliament and of the
  Council Laying Down Harmonised Rules on Artificial Intelligence (Artificial
  Intelligence Act) and Amending Certain Union Legislative Acts}, 2021.
\newblock COM/2021/206 final.

\bibitem[Gel20]{RG2020}
Rapha\"{e}l Gellert.
\newblock {\em The Risk-Based Approach to Data Protection}.
\newblock Oxford University Press, first edition, December 2020.

\bibitem[Gel21]{RG2021}
Rapha\"{e}l Gellert.
\newblock The role of the risk-based approach in the general data protection
  regulation and in the european commission’s proposed artificial
  intelligence act: Business as usual?
\newblock {\em Journal of Ethics and Legal Technologies}, 3(2):15--33, November
  2021.

\bibitem[GG13]{GG2013}
Rapha\"{e}l Gellert and Serge Gurtwirth.
\newblock The legal construction of privacy and data protection.
\newblock {\em Computer Law \& Security Review}, 29(5):522--530, 2013.

\bibitem[{IEE}19]{IEEE2019}
{IEEE Global Initiative on Ethics of Autonomous and Intelligent Systems}.
\newblock Ethically aligned design: A vision for prioritizing human well-being
  with autonomous and intelligent systems.
\newblock Technical report, Manhattan, New York, 2019.
\newblock First Edition.

\bibitem[IH18]{Isa_han:j:Privacy}
Jim Isaak and Mina~J. Hanna.
\newblock User data privacy: Facebook, cambridge analytica, and privacy
  protection.
\newblock {\em Computer}, 51(8):56--59, 2018.

\bibitem[{Int}13]{ISO27001}
{International Organization for Standardization}.
\newblock {Information technology -- Security techniques -- Information
  security management systems -- Requirements}, 2013.
\newblock ISO/IEC 27001:2013.

\bibitem[{Int}19a]{ISO27018}
{International Organization for Standardization}.
\newblock {Information technology -- Security techniques -- Code of practice
  for protection of personally identifiable information (PII) in public clouds
  acting as PII processors}, 2019.
\newblock ISO/IEC 27018.

\bibitem[{Int}19b]{ISO27701}
{International Organization for Standardization}.
\newblock {Security techniques — Extension to ISO/IEC 27001 and ISO/IEC 27002
  for privacy information management — Requirements and guidelines}, 2019.
\newblock ISO/IEC 27701:2019.

\bibitem[{Int}21]{ISO27002}
{International Organization for Standardization}.
\newblock {Information security, cybersecurity and privacy protection --
  Information security controls}, 2021.
\newblock ISO/IEC DIS 27002.

\bibitem[JWHT21]{JWHT-stat-learning}
Gareth James, Daniela Witten, Trevor Hastie, and Robert Tibshirani.
\newblock {\em {An Introduction to Statistical Learning with Applications in
  R}}.
\newblock Springer, 2021.

\bibitem[KL13]{KoopsLeenes2013_Privacy}
Bert-Jaap Koops and Ronald Leenes.
\newblock Privacy regulation cannot be hardcoded. a critical comment on the
  ‘privacy by design’ provision in data-protection law.
\newblock {\em International Review of Law, Computers \& Technology},
  28(2):159--171, 2013.

\bibitem[MKGCM14]{NEO-PRI}
Jessica Maples-Keller, Li~Guan, Nathan Carter, and Joshua Miller.
\newblock A test of the international personality item pool representation of
  the revised neo personality inventory and development of a 120-item
  ipip-based measure of the five-factor model.
\newblock {\em Psychological assessment}, 26, 06 2014.

\bibitem[Mok18]{DM2018}
Dorota Mokrosinska.
\newblock Privacy and autonomy: On some misconceptions concerning the political
  dimensions of privacy.
\newblock {\em Law and Philosophy}, 37(2):117--143, 2018.

\bibitem[Nis09]{HN2009}
Helen Nissenbaum.
\newblock {\em Privacy in Context: Technology, Policy, and the Integrity of
  Social Life}.
\newblock Stanford University Press, first edition, November 2009.

\bibitem[Nis18]{HN2018}
Helen Nissenbaum.
\newblock Respecting context to protect privacy: Why meaning matters.
\newblock {\em Science and Engineering Ethics}, 24(4):831--852, 2018.

\bibitem[NM65]{NelderMead1965_minimization}
J.A. Nelder and R.~Mead.
\newblock A simplex method for function minimization.
\newblock {\em The Computer Journal}, 7:308--313, 1965.

\bibitem[{Olm}28]{OvUS1928}
{Olmstead v United States}.
\newblock {277 U.S. 438}, 1928.

\bibitem[{Org}13]{OECD2013}
{Organisation for Economic Co-Operation and Development}.
\newblock {The OECD Privacy Framework}, 2013.
\newblock https://www.oecd.org/sti/ieconomy/oecd\_privacy\_framework.pdf.

\bibitem[{Org}19]{OECD2019}
{Organisation for Economic Co-operation and Development}.
\newblock {Recommendation of the Council on Artificial Intelligence}.
\newblock
  \url{https://legalinstruments.oecd.org/en/instruments/OECD-LEGAL-0449}, 2019.
\newblock Adopted on 22 May 2019, OECD-LEGAL-0449.

\bibitem[Rul19]{JR2019}
James~B. Rule.
\newblock Contextual integrity and its discontents: A critique of helen
  nissenbaum's normative arguments.
\newblock {\em Policy and Internet}, 11(3):260--279, August 2019.

\bibitem[{The}12]{PBR2012}
{The White House}.
\newblock Consumer data privacy in a networked world: A framework for
  protecting privacy and promoting innovation in the global digital economy,
  February 2012.
\newblock https://obamawhitehouse.archives.gov/sites/default/files/pri
  vacy-final.pdf.

\bibitem[VGO{\etalchar{+}}19]{scipy}
Pauli {Virtanen}, Ralf {Gommers}, Travis~E. {Oliphant}, Matt {Haberland}, Tyler
  {Reddy}, David {Cournapeau}, Evgeni {Burovski}, Pearu {Peterson}, Warren
  {Weckesser}, Jonathan {Bright}, St{\'e}fan~J. {van der Walt}, Matthew
  {Brett}, Joshua {Wilson}, K.~{Jarrod Millman}, Nikolay {Mayorov}, Andrew
  R.~J. {Nelson}, Eric {Jones}, Robert {Kern}, Eric {Larson}, CJ~{Carey},
  {\.I}lhan {Polat}, Yu~{Feng}, Eric~W. {Moore}, Jake {Vand erPlas}, Denis
  {Laxalde}, Josef {Perktold}, Robert {Cimrman}, Ian {Henriksen}, E.~A.
  {Quintero}, Charles~R {Harris}, Anne~M. {Archibald}, Ant{\^o}nio~H.
  {Ribeiro}, Fabian {Pedregosa}, Paul {van Mulbregt}, and SciPy 1.~0
  {Contributors}.
\newblock {SciPy 1.0--Fundamental Algorithms for Scientific Computing in
  Python}.
\newblock {\em arXiv e-prints}, page arXiv:1907.10121, Jul 2019.

\bibitem[WB90]{WB1890}
Samuel~D. Warren and Louis~D. Brandeis.
\newblock The right to privacy.
\newblock {\em Harvard Law Review}, 4(5):193--220, December 1890.

\bibitem[Wes67]{AW1967}
Alan~F. Westin.
\newblock {\em Privacy and Freedom}.
\newblock London: The Bodley Head, first edition, November 1967.

\bibitem[Zee76]{Zeeman1976_catastrophe}
E.~C. Zeeman.
\newblock Catastrophe theory.
\newblock {\em Scientific American}, 234(4):65--83, 1976.

\end{thebibliography}
\bibliographystyle{alpha}
}

\end{document}